\documentclass[aps,prb,showpacs,preprint,amsfonts,amssymb,amsmath]{revtex4}
\usepackage{graphicx}
\usepackage{amsmath}
\usepackage{bm}

\begin{document}

\title{The focusing of electron flow in a bipolar Graphene ribbon
with different chiralities}

\author{Yanxia Xing$^1$, Jian Wang$^{1,\ast}$,
and Qing-feng Sun$^{2}$}

\address{
$^1$Department of Physics and the Center of Theoretical and
Computational Physics, The University of Hong Kong, Pokfulam Road,
Hong Kong, China\\ $^2$Beijing National Lab for Condensed Matter
Physics and Institute of Physics, Chinese Academy of Sciences,
Beijing 100190, China}

\begin{abstract}
The focusing of electron flow in a symmetric p-n junction (PNJ) of
graphene ribbon with different chiralities is studied. Considering
the PNJ with the sharp interface, in a armchair ribbon, the electron
flow emitting from $(-L,0)$ in n-region can always be focused
perfectly at $(L,0)$ in p-region in the whole Dirac fermion regime,
i.e. in whole regime $E_0<t$ where $E_0$ is the distance between
Dirac-point energy and Fermi energy and $t$ is the nearest hopping
energy. For the bipolar ribbon with zigzag edge, however, the
incoming electron flow in n-region is perfectly converged in
p-region only in a very low energy regime with $E_0<0.05t$.
Moreover, for a smooth PNJ, electrons are backscattered near PNJ,
which weakens the focusing effect. But the focusing pattern still
remains the same as that of the sharp PNJ. In addition, quantum
oscillation in charge density occurs due to the interference between
forward and backward scattering. Finally, in the presence of weak
perpendicular magnetic field, charge carriers are deflected in
opposite directions in the p-region and n-region. As a result, the
focusing effect is smeared. The lower energy $E_0$, the easier the
focusing effect is destroyed. For the high energy $E_0$ (e.g.
$E_0=0.9t$), however, the focusing effect can still survive in a
moderate magnetic field on order of one Tesla.
\end{abstract}

 \maketitle

\section{introduction}
Graphene is a single layer carbon atoms packed into honeycomb
lattice. From the point of view of its electronic properties in the
low energy regime, a graphene sheet is a two-dimensional (2D)
zero-gap semiconductor with the conical energy spectrum around Dirac
points, the corners of the hexagonal first Brillouin zone, and its
quasi-particles are formally described by the massless Dirac
equation where the speed of light is replaced by the Fermi velocity
of graphene.\cite{Dirac} The detailed electronic properties of
graphene has been reviewed in Ref.[\onlinecite{mod1}]. Different
from the usual zero-gap semiconductor in which the electrons and
holes are normally described by separate Schr\"{o}dinger equations
with generally different effective masses, the electrons and holes
in graphene are conjugately linked and described by different
components of the same spinor wavefunction,\cite{Dirac} which means
they are interconnected as Dirac fermions in QED. So graphene is a
relativistic counterpart in the condensed-matter system. So far, 2D
graphene has been successfully fabricated
experimentally.\cite{ref1,ref01} By varying the gate
voltage\cite{ref3} or doping the underlying substrate,\cite{ref4}
the charge carriers of graphene can be easily tuned, the
controllable ballistic PNJ or PNPJ are also realized
experimentally.\cite{pnj} Therefore intriguing phenomena exhibited
in the bipolar
graphene,\cite{microwave,Andreev,aref1,Klein,Kleinback,lens,Kleininter,aref2}
such as microwave-induced reflection,\cite{microwave} specular
Andreev reflection,\cite{Andreev,aref1} Klein tunneling,\cite{Klein}
Klein backscattering\cite{Kleinback} and negative refraction index
effect\cite{lens}, are possible to be verified experimentally. In
fact, a direct experimental observation of Klein tunneling has been
realized through a extremely sharp graphene PNJ.\cite{Kleininter}

It was shown that due to the Berry phase $\pi$, which was derived
from the intersection of the energy bands at Dirac
points,\cite{Berry} the backscattering is absent.\cite{Ando1} This
naturally leads to the so-called Klein tunneling\cite{Klein} or
interband tunneling that an incident electron tunnel from the
conduction into the valence band without backscattering. Because of
the conservation of momentum and energy, interband tunneling through
the p-n interface may resemble the optical refraction at the surface
of meta-materials with negative refractive
index.\cite{negrefconcept} In another word, the Klein paradox gives
rise to the negative refraction.\cite{negtolens} This means that an
interface of the symmetric PNJ perpendicular to the current flow is
able to focus the electric current whereas a ballistic strip of
p-type graphene separated by two n-type regions acts as a lens.
These intriguing phenomena have been described in
Ref.[\onlinecite{lens}], in which the Kubo formula was applied to
the single-particle Dirac-like Hamiltonian of graphene. It means
that for an infinite 2D graphene system with ideal conical energy
spectrum, i.e., in the very low energy regime, the electrons
emitting from source are perfectly focused at the mirror symmetric
point of the symmetric PNJ. For the realistic graphene system,
however, one can not separate electrons and holes close to Dirac
point due to the electron-hole puddles\cite{puddle} (about tens
$meV$\cite{puddle1}). Of course, we can experimentally increase the
density of electrons and holes by the gate voltage to evade from the
puddle region. However, in the high density case, the energy
spectrum deviates from the linear relation. Hence, the effect due to
the non-linear dispersion should be examined. For this purpose, we
will use the tight binding Hamiltonian to study the graphene based
PNJ. In addition, considering the chirality of graphene ribbon, it
is better to use the tight-binding Hamiltonian to describe the
transport processes along different chirality directions.

In this paper, using the tight-binding model, we carry out a
theoretical study on the focusing effect of electron flow in the
graphene ribbon with a symmetric PNJ. Due to the chirality of
graphene, the focusing effects may be different for the zigzag
ribbon and the armchair ribbon. Indeed, it is found that for the
armchair ribbon with a sharp p-n interface the electron flow
emitting from $(-L,0)$ in n-region can always be focused perfectly
at $(L,0)$ in p-region for all energy $E_0=|E_F-E_{p/n}|<t$. For the
zigzag ribbon, however, the electron flow is perfectly focused only
in the very low regime ($E_0<0.05t$). Furthermore, the perfect
focusing in the bipolar ribbon is robust against disorders induced
by the random potential. But the edge disorders drastically affect
the perfect focusing. For a smooth PNJ, electrons are backscattered
close to PNJ at a distance proportional to $k_y$ and interface width
$d$.\cite{mod} In this case a quantum interference between forward
and backward scattering is present and the intensity of the focused
spot is weakened, but the focusing pattern keeps almost. Finally, in
the presence of a weak perpendicular magnetic field, the momentum
$k_y$ is no longer conserved. Consequently particles are deflected
in opposite directions in the p-region and n-region, which destroys
the perfect focusing especially in the low energy regime.

The rest of the paper is organized as follows. In Sec. II, the model
system including bipolar graphene ribbon in the tight-binding
representation with attached source or detector terminal is
introduced. The formalisms for calculating the local particle
density, the local current density vector and the local conductance
are then derived. Sec. III gives numerical results along with some
discussions. Finally, a brief summary is presented in Sec. IV.

\section{model and formalism}

In order to study the scattering due to PNJ, we consider two kinds
of open bipolar graphene systems (armchair and zigzag ribbons) as
shown in Fig.1. The bipolar graphene ribbon consists of
semi-infinite electronlike ribbon [orange region] and semi-infinite
holelike ribbon [green region] along $x$-direction with a sharp p-n
interface located at $x=0$. Electron flow is injected into graphene
system from a source lead located at $(-L,0)$ in the n-region. Here
we assume that the source lead and the bipolar ribbon are in contact
with six lattices [see the blue area in Fig.1]. The injected
electrons in the n-region can spread in all directions. Because of
the open boundary condition, left-going electrons can finally escape
into infinite graphene electrode while right-going electrons [shown
in Fig.2] can then be scattered only by p-n interface [thick black
line]. Consequently, the response signals are converged around the
symmetric site $(L,0)$ [red area in Fig.1]. In order to investigate
the focusing current, we couple a detecting electrode locally in the
p-region and study the local current (conductance) flowing from that
electrode. Clearly, the local current depends on the coupling
position of the detecting electrode.

The total Hamiltonian including the infinite graphene ribbon in the
tight-binding representation\cite{ham} and the source/drain
electrode that is expressed in $k$ space with the free electron
model can be written as:
\begin{eqnarray}
 H &=&\sum_{\bf i} \epsilon_{\bf i} a^{\dagger}_{\bf i} a_{\bf i}
      -\sum_{<{\bf ij}>} t e^{i\phi_{\bf ij}} a_{\bf i}^{\dagger} a_{\bf
      j}\nonumber \\
      &+&\sum\limits_{\alpha,k} \left[ \epsilon_{\alpha,k} d_{\alpha,k}^\dagger
d_{{\alpha},k} + (\gamma a_{{\bf i}_{\alpha}}^\dagger d_{{\alpha},k}
+ h.c)\right] \label{Ham}
\end{eqnarray}
where ${\bf i}=({\bf i}_x , {\bf i}_y)$ is the index of the discrete
site on the honeycomb lattice which is sketched in the Fig.1, and
$a_{\bf i}$ and $a_{\bf i}^{\dagger}$ are the annihilation and
creation operators at the site ${\bf i}$. Here $\epsilon_{\bf i}$ in
the first term of Eq.(1) is the on-site energy (i.e., the energy of
the Dirac point) which can be controlled experimentally by the gate
voltage. In the n-region or p-region far away from PNJ all the
on-site energy are the same with $\epsilon_{{\bf i}}=E_n$ or
$\epsilon_{{\bf i}}=E_p$. Near PNJ, $\epsilon_{\bf i}$ changes from
$E_n$ to $E_p$ abruptly or smoothly for the sharp PNJ or smooth PNJ,
respectively. The second term in Eq.(1) is the nearest neighbor
hopping term with the hopping energy $t$, `$<{\bf ij}>$' denotes the
nearest neighbor lattice sites. When the graphene ribbon is under a
uniform perpendicular magnetic field $B_z=B$, a phase $\phi_{\bf
ij}$ is added in the hopping term, and $\phi_{\bf ij}=\int_{\bf
i}^{\bf j} \vec{A} \cdot d\vec{l}/\phi_0$ with the vector potential
$\vec{A}=(-By,0,0)$ and the flux quanta $\phi_0=\hbar/e$. Finally,
the last term in Eq.(1) represents the Hamiltonian of the source and
detector leads described in the $k$ space and their coupling to the
graphene lattices ${\bf i}_{\alpha}$. Here $\alpha=s,d$ represent
source and detecting electrodes and $d_{{\alpha},k}$
($d^\dagger_{{\alpha},k}$) is the annihilation (creation) operator
of the electrons in the electrode $\alpha$.

When the electron flow is injected from the source electrode into
the graphene in the n-region, the response signal is induced
everywhere in the p-region. To make a thorough study on the focusing
effect, we consider three physical quantities in the p-region: (1)
the local current density vector, (2) the local particle density,
and (3) the local current (conductance) of the detecting electrode.
For quantities (1) and (2), we consider the system without the
coupling of the detecting electrode so that the influence of the
detecting electrode can be eliminated.

\subsection{Local current density vector}
The general current density vector
$\mathbf{J}_{\mathbf{i}\mathbf{j}}$ from the site ${\bf i}$ to its
nearest neighbor site ${\bf j}$ can be expressed as:\cite{DOC}
\begin{eqnarray}
J_{\mathbf{i}\mathbf{j}}&=&\frac{e}{h}\int dE~~\left[ {\bf
G}^<_{\mathbf{i}\mathbf{j}}(E){\bf H}_{\mathbf{j}~\mathbf{i}}- {\bf
H}_{\mathbf{i}\mathbf{j}}{\bf G}^<_{\mathbf{j}~\mathbf{i}}(E)
\right]\nonumber \\
&=&2\frac{e}{h}{\rm Re} \int dE~~\left[ te^{i\phi_{\bf ji}}{\bf
G}^<_{\mathbf{i}\mathbf{j}}(E)\right]\label{DOC0}
\end{eqnarray}
where $e$ is the electron charge, ${\bf G}^<_{\bf ij}$ is the matrix
element of the lesser Green's function of the scattering region.
Because the graphene ribbon is translation invariant in the
p/n-region, the central scattering region can be chosen arbitrarily
as long as the source sites ${\bf i}_s$ and the detector sites ${\bf
i}_d$ are included. From the Keldysh equation, the lesser Green's
function is related to the retarded and advanced Green's functions,
\begin{eqnarray}
{\bf G}^<(E)={\bf
G}^r(E)\left[\sum_\alpha{\bf\Sigma}^<_\alpha(E)\right]{\bf G}^a(E)
\end{eqnarray}
Here the sum index $\alpha=L,R,s$ denote the left/right graphene
lead and source lead with $\alpha\neq d$ because of the decoupling
of the detector lead. The retarded Green's function ${\bf
G}^r(E)=[{\bf G}^a(E)]^{\dagger}=\{E{\bf I}-{\bf
H}_0-\sum_{\alpha}{\bf\Sigma}^r_{\alpha}(E)\}^{-1}$, where ${\bf
H}_0$ is Hamiltonian matrix of the central scattering region and
${\bf I}$ is the unit matrix with the same dimension as that of
${\bf H}_0$, ${\bf \Sigma}_{\alpha}^r$ is the retarded self-energy
function from the lead-$\alpha$. ${\bf\Sigma}^r_\alpha$ can be
obtained from ${\bf\Sigma}^r_{L/R}(E)={\bf H}_{c,L/R}{\bf
g}^r_{L/R}(E){\bf H}_{L/R,c}$, where ${\bf H}_{c,L/R}$ (${\bf
H}_{L/R,c}$) is the coupling from central region (lead-L/R) to
lead-L/R (central region) and ${\bf g}^r_{L/R}(E)$ is the surface
retarded Green's function of the semi-infinite lead which can be
calculated using a transfer matrix method.\cite{transfer} Concerning
the source lead, we take the wide band approximation, then the
non-zero elements of self energy matrix
${\bf\Sigma}^r_s(E)=-i{\bf\Gamma}_s/2$ is energy independent, where
line-width function ${\bf\Gamma}_s=2\pi \gamma^2\rho_s(E_F)$. ${\bf
\Sigma}^<_\alpha(E)$ in Eq.(3) is the lesser self energy of the
lead-$\alpha$. Because the isolated lead is in the equilibrium,
${\bf\Sigma}^<_\alpha$ can be obtained from the
fluctuation-dissipation theorem:
\begin{eqnarray}
{\bf\Sigma}^<_\alpha(E)&=&\left[{\bf\Sigma}^a_\alpha(E)-{\bf\Sigma}^r_\alpha(E)\right]f_\alpha(E)\nonumber
\\
&=&i{\bf\Gamma}_\alpha(E)f_\alpha(E)
\end{eqnarray}
with ${\bf\Sigma}^a_\alpha={\bf\Sigma}^{r,\dagger}_\alpha$ and
$f_\alpha(E)=f_0(E-eV_\alpha)$ where $f_0(E) =1/[{\rm
exp}(E/k_B\mathcal{T})+1]$ is the Fermi distribution function.
$V_\alpha$ is the external bias in the terminal-$\alpha$. Since we
are interested only in the local response due to the source flow,
the external biases are set as $V_s=\delta V$ and $V_{L/R}=0$. In
calculating transport properties, we divide $G^<(E)$ into
equilibrium and non-equilibrium parts as
\begin{eqnarray}
{\bf G}^<(E)&=&{\bf G}^r(E)\left[if_0(E)\sum_\alpha{\bf\Gamma}_\alpha(E)\right]{\bf G}^a(E)\nonumber \\
&+& {\bf
G}^r(E)\left[i\sum_\alpha\left\{f_\alpha(E)-f_0(E)\right\}{\bf\Gamma}_\alpha(E)\right]{\bf
G}^a(E)
\end{eqnarray}
where the equilibrium term does not contribute to the transport and
can be dropped out from now on. It is the non-equilibrium term that
gives rise to the system response to the electron injection from the
source lead. Because of $V_s=\delta V$ and $V_{L/R}=0$, we have
\begin{eqnarray}
{\bf G}^<(E)=i{\bf
G}^r(E)\left[f_s(E)-f_0(E)\right]{\bf\Gamma}_s{\bf
G}^a(E)\label{Gless}
\end{eqnarray}
Substituting Eq.(6) into Eq.(2) and considering the limit of small
source bias, the local current density vector ${\bf J}_{\bf ij}$ (or
the local conductance density vector ${\bf J}_{\bf ij}/\delta V$)
can be expressed in the following form at zero temperature:
\begin{equation}
{\bf J}_{\mathbf{i}\mathbf{j}}/\delta V =\frac{2e}{h}{\rm Im}
\left\{ te^{i\phi_{\bf ji}}\left[{\bf G}^r(E_F)i{\bf\Gamma}_s{\bf
G}^a(E_F)\right]_{\bf ij}\right\} \label{DOC1}
\end{equation}
It should be noted that the current density ${\bf J}_{\bf ij}$ in
Eq.(\ref{DOC1}) is defined between the lattice sites ${\bf i}$ and
${\bf j}$ with the direction from site ${\bf i}$ to ${\bf j}$. In
order to obtain the local current density vector ${\bf J_i}$ at the
site ${\bf i}$, we take the weighted average on ${\bf J}_{\bf ij}$
over all the nearest neighbors ${\bf j}$.

\subsection{Local particle density}
The local particle density (i.e. the electron occupation number) is
defined as
\begin{eqnarray}
\rho_{\bf i}&=&-ie\int \frac{dE}{2\pi} ~~{\bf G}^<_{\bf
ii}(E)\label{rho}
\end{eqnarray}
where ${\bf G}^<_{\bf ii}$ is the diagonal element of the lesser
Green's function ${\bf G}^<$ in Eq.(3). Similar to the derivation of
local current density vector ${\bf J}_{\bf ij}/\delta V$, here we
consider only the variation of the local particle density caused by
the electron injection from the source lead. At zero temperature and
the small bias $\delta V$ limit, substituting Eq.(6) into Eq.(8),
the variation of the local particle density is expressed as:
\begin{eqnarray}
\delta\rho_{\bf i}/\delta V&\equiv&
 [\rho_{\bf i}(V_s=\delta V)-\rho_{\bf i}(V_s=0)]/\delta V
 \nonumber\\
 &=& \frac{e^2}{2\pi} \left[{\bf G}^r(E_F){\bf \Gamma}_s{\bf G}^a(E_F)\right]_{\bf ii}
\end{eqnarray}

Since the Hamiltonian is defined at discrete lattice sites, the
local quantities can also be defined at each lattice site. Such a
local quantity is feasible but not necessary. In fact, for graphene,
we can define the ``local" quantity by averaging over six discrete
sites in a unit cell of honeycomb lattice. This average can
eliminate the strong variation of local quantities in the A and B
sublattices. Now every local site can be determined from the
coordinates $(x,y)$ shown in Fig.1. For example, the local injection
area displayed in a blue area in Fig.1 is located at $(-3,0)$ in
Fig.1(a) and $(-5,0)$ in Fig.1(b). In the whole lattice region in
Fig.1, there are $7\times 3$ and $11\times 3$ units in Fig.1(a) and
Fig.1(b), respectively.

\subsection{Local conductance}
Concerning the local conductance, the detecting lead-d is coupled to
the graphene ribbon in the p-region. Similar to the local particle
density, here the detecting lead also couples to six sites ${\bf
i}_d$ in a unit cell of honeycomb lattice. The current flowing to
the detecting lead-$d$ can be expressed as
\begin{eqnarray}
 J_{d}&=&\frac{e}{\hbar}\sum_{k_d}\left[ {\bf G}^<_{{\bf
i}_d,k_d}(t_1,t_2){\bf H}_{k_d,{\bf i}_d}- {\bf H}_{{\bf
i}_d,k_d}{\bf G}^<_{k_d,{\bf
i}_d}(t_1,t_2)\right]_{t_1=t_2}\nonumber \\
 &=&\frac{e}{\hbar}\sum_{k_d}\left] {\bf G}^<_{{\bf i}_d,k_d}(t_1,t_2)\gamma- \gamma
 {\bf G}^<_{k_d,{\bf i}_d}(t_1,t_2)\right]_{t_1=t_2} \label{Cur1}
\end{eqnarray}
Using the Dyson equation in the time contour, we can get the
Landauer-B$\ddot{u}$ttiker formula\cite{Landuer} which is expressed
in terms of non-equilibrium Green's functions:
\begin{eqnarray}
 J_{d}&=&\frac{e}{\hbar}\sum_{\alpha} \int\frac{dE}{2\pi}
{\bf T}_{d,\alpha}(E)[f_{d}(E)-f_{\alpha}(E) ].
\end{eqnarray}
with $\alpha=L,R,s$ representing the left/right graphene leads and
source leads. Since we shall concentrate only on the response
current induced by the current injected from the source lead, we use
the following boundary conditions $V_{L,R}=V_d=0$ and $V_s=\delta
V$. The current now becomes,
\begin{eqnarray}
 J_{d}&=&\frac{e}{\hbar} \int\frac{dE}{2\pi}
[ T_{d,s}(E)(f_{d}(E)-f_{s}(E)) ].
\end{eqnarray}
where $T_{d,s}$ is the transmission coefficient from the source lead
located at the site ${\bf i}_s$ to the detecting lead located at the
site ${\bf i}_d$ which can be calculated from $T_{d,s}(E)={\rm
Tr}\left[{\bf \Gamma}_{d}{\bf G}^r(E){\bf\Gamma}_{s}{\bf
G}^a(E)\right]$, where ${\bf G}^{a}={\bf G}^{r\dagger}$ is the
advanced Green's function in the scattering region. In the wide band
limit, the line-width function
${\bf\Gamma}_{s/d}(E)=i({\bf\Sigma}_{s/d}^r-{\bf\Sigma}_{s/d}^{r\dagger})=2\pi
\gamma^2\rho_{s/d}(E_F)$. Here $f_{s/d}(E)$ in Eq.(12) is the Fermi
distribution function of the source and detecting lead, and
$f_s(E)=f_0(E-eV_s)$ and $f_d(E)=f_0(E)$. Considering the zero
temperature and small bias $V_{s}$ limits, local conductance
contributed by the source electron flow can be expressed as:
\begin{eqnarray}
 G_{{\bf i}_d}=J_d/\delta V
 =\frac{e^2}{h} T_{d,s}(E_F).
\end{eqnarray}

\section{numerical results and discussion}

In the numerical calculations, we set the nearest-neighbor
carbon-carbon distance $a=0.142nm$ and the second nearest-neighbor
distance $b=\sqrt3a\simeq0.25nm$, the hopping energy $t=2.75eV$ as
in a real graphene sample.\cite{ref3} In this paper, we consider
only the focusing effect of the symmetric PNJ, in which the electron
density of n region is the same as the hole density in p region,
i.e., $\rho_e=\rho_h$. For simplicity, we set $E_F=0$. Hence in the
n-region far away from PNJ the on site energy $\epsilon_{{\bf
i}}=E_n=-E_0$ while $\epsilon_{{\bf i}}=E_p=E_0$ in the p-region far
away from PNJ. Near PNJ, $\epsilon_{\bf i}$ changes from $-E_0$ to
$E_0$ abruptly (smoothly) for the sharp (smooth) PNJ.

Experimentally, it is more convenient to measure the electric
conductance. So, in order to detect the focusing effect by a single
PNJ in graphene, one can use a small electric contact (such as a STM
probe) as a source of electron flow in the n-region and another
local probe located in the p-region as a detector. Electric
conductance between the two contacts measures the transmission
probability for a charged carrier from the source to the detector.
Numerically, We have calculated the local conductance $G_{{\bf
i}_d}$ and confirmed that the distribution of local conductance is
similar to that of the local particle density $\delta\rho_{\bf
i}/\delta V$. For this reason, only the numerical results on local
particle density are shown in this paper. In addition, in order to
visualize the focusing process, we also show the distribution of
local current density vector in the p-region.

\subsection{Focusing effect in very low energy regime}
Now we study the focusing effect in the graphene ribbon with a sharp
PNJ. For a zigzag ribbon or a armchair ribbon with sharp and
symmetric PNJ, the spacial distribution of the local particle
density $\delta \rho(x,y)/\delta V$ in p-region due to electrons
coming from the source lead is shown in Fig.3(a) and Fig.3(b),
respectively. Following observations are in order. First of all,
electrons injected at $(-L,0)$ in n-region can be focused around
$(L,0)$ shown as red spot in Fig.3 which is similar to
Ref.[\onlinecite{lens}]. This is because the Fermi energy $E_F$ is
close to Dirac energy $E_0= 0.05t$ so that the energy dispersion is
nearly linear, i.e., $E_0\simeq kb\frac{\sqrt{3}}{2}t$ where $k$ is
module of momentum vector $\mathbf{k}$. The charged carriers
scattering through PNJ can mimic the refraction of light by
left-handed metamaterials with refraction index equal to -1.
Secondly, besides the focusing spot (red and green region), there is
also a weak interference pattern (blue wave pattern) shown in Fig.3,
which is different from Ref.[\onlinecite{lens}] in which the wave
pattern is absent. In fact, the wave pattern is solely due to the
boundary of the nanoribbon. When an electron is injected from the
source area, it can propagate in all directions and the right-going
electrons can be scattered by either the boundary of nanoribbon
(thin black lines) or sharp PNJ (thick black line) as shown in
Fig.2. In the p-region, interference pattern is due to the
interference between the state scattered by both boundary and PNJ
(thick blue lines) and the state scattered only by PNJ (thick red
lines). The spacial period of the interference is proportional to
the momentum ${\mathbf k}$ or $E_0$. Finally, the focusing phenomena
in zigzag ribbon is slightly different from that of armchair ribbon:
the electron flow is perfectly focused in armchair ribbon [panel
(b)], but can't be fully focused in the zigzag ribbon [panel (a)].
This is because the energy band structures are different for the
armchair ribbon and zigzag ribbon. In the following, we will examine
the different focusing effects in detail.

\subsection{Focusing in zigzag ribbon}

When Fermi energy is gradually moved away from Dirac point, the
energy spectrum is not ideal conical anymore. In Fig.4 we plot the
contour lines of dispersion relation $E(k_x,k_y)$ of graphene
sheet\cite{footnote} with energy interval between nearest contour
lines $\delta E=0.1t$. The panel (a) is for the graphene sheet with
the carbon-carbon bond along the $x$-direction which corresponds to
the armchair graphene ribbon, and the panel (b) is for the graphene
sheet with the carbon-carbon bond along the $y$-direction
corresponding to the zigzag ribbon. The deviation of ideal conical
energy spectrum is clearly exhibited even at small energy $E=0.2t$
(the second small contour lines around the Dirac points $K$ and $K'$
show anisotropy behaviors). Since the p-n interface at $x=0$ is
along $y$ direction, the $y$ component of momentum, $k_y$, is
conserved during the scattering. As a result, the incident wave
vector $k_{x,in}$, the reflecting wave vector $k_{x,r}$, and the
transmitting wave vector $k_{x,t}$ must lie on the black dotted
lines in Fig.4.
%
%For there are three equivalent $K$ and $K'$ corners in the first
%Brillouin zone, we chose $K=(0,\frac{2}{3}\frac{2\pi}{b})$ for
%panel(a) and
%$K=(\frac{2}{3}\frac{2\pi}{b},0),K'=(-\frac{2}{3}\frac{2\pi}{b},0)$
%for panel (b) to study the relation between injection momentum
%$k_{in}$ and reflection/transmiting momentum $k_{r/t}$.
When an electron with energy $E=E_F$, velocity $(v_x,v_y)$ and
corresponding momentum $(k_{x,in},k_{y,in})$ with respect to Dirac
point $K$, injects from n-region and is scattered at the p-n
interface, according to the identical direction of $V_x$, we can
solve the reflecting and scattering momentum $k_{x,r}$ and $k_{x,t}$
using the energy conservation and $k_y$ conservation. For the zigzag
ribbon [corresponding to Fig.4(b)], $k_{x,in}$ can be
intra-scattered to $k_{x,r/t}$ around $K$
[$K=(\frac{2}{3}\frac{2\pi}{b},0)$] valley or inter-scattered to
$k_{x',r/t}$ in $K'$ [$K'=-(\frac{2}{3}\frac{2\pi}{b},0)$] valley.
The interband scattering states is symmetric which satisfies
$k'_{x,r/t}=-k_{x,in}$. The intraband scattering, however, exhibits
asymmetric properties that $\delta k_{x,r/t} \equiv
k_{x,in}+k_{x,r/t}\ne 0$ or $v_{x,in}\ne -v_{x,r}$, $v_{x,in}\ne
v_{x,t}$ for any fixed $k_y$. In Fig.5, we plot the asymmetric
relation between $k_{x,in}$ and $k^{intra}_{x,r/t}$ in the intraband
scattering and the symmetric relation between $k_{x,in}$ and
$k^{inter}_{x,r/t}$ in the interband scattering. We see that in the
intraband scattering, the larger $k_y$, the larger derivation
$\delta k^{intra}_{x,r/t}$ is, while in the interband scattering,
$k^{inter}_{x,r/t}$ is always equal to $-k_{x,in}$ for all $k_y$. It
is known that interband scattering is weak\cite{valleysuppress} in
pure samples due to the large momentum shift, so the asymmetric
intraband scattering is dominant in zigzag ribbon PNJ. As a result,
the refraction index can not be strictly equal to $-1$ and the
charge flow can't be fully converged at the symmetric spot. In
Fig.\ref{zig} focusing effect for $E_0=0.1t$ and $E_0=0.2t$ are
plotted, respectively. Comparing Fig.\ref{zig}(a) and
Fig.\ref{zig}(b) we see that it is more difficult to focus the
electron beam for larger momentum ${\mathbf k}$ (energy $E_0$).

\subsection{Focusing in armchair ribbon}

For the armchair ribbon [corresponding to Fig.4(a)] only intraband
scattering occurs. Now $k_{x,in}$ can be symmetrically
intra-scattered to $k_{x,r/t}$ in $K$ valley with
$-k_{x,in}=k_{x,r}=k_{x,t}$ since $k_x$ in Fig.4(a) is symmetric
about $k_x=0$. So, although the energy dispersion of armchair ribbon
is also not strictly linear at high energies as in zigzag ribbon,
due to the symmetric scattering, the focusing effect is always
perfect in armchair ribbon for all $E_0<t$ (i.e. in Dirac fermion
regime). Furthermore, with increasing of $E_0$, the electron flow
coming from n-region shows better convergence in the p-region with
smaller focusing spot and stronger intensity. Furthermore, The
spacial period of the interference pattern is proportional to the
momentum ${\mathbf k}$ or $E_0$, which can be clearly seen by
comparing Fig.3(b), Fig.\ref{arm1}(a) and Fig.\ref{arm1}(b). .

Roughly speaking, two energy regimes are considered for the Dirac
Fermion according to band structure of graphene: (1). 'Near linear
dispersion' regime $0<E_0<0.5t$ where $E_0\approx
kb\frac{\sqrt{3}}{2}t$. (2) 'Beyond linear dispersion' regime
$0.5t<E_0<t$ where the energy spectrum is non-conical. The focusing
effect corresponding to these two regimes are plotted in
Fig.\ref{arm1} and Fig.\ref{arm2}, respectively. In the first
regime, with the near linear dispersion relation, velocity $v_x$ or
$v_y$ is roughly a constant, and ${k_{y,in}}/{k_{x,in}}\approx
{v_{y,in}}/{v_{x,in}}$, ${k_{y,t}}/{k_{x,t}}\approx
{v_{y,t}}/{v_{x,t}}$.
For the symmetric scattering ($k_{x,in}=k_{x,t}$) in the armchair
ribbon, refraction index $n\approx -1$ giving rise to the convergent
spot shown in Fig.\ref{arm1}. When $E_0$ is large enough (larger
than 0.5t), energy spectrum is non-conical and velocity now depends
on momentum. This leads to a different focusing effect shown in
Fig.\ref{arm2} in which crossed focusing zone is present.

In order to show the focusing of the electron flow vividly, instead
of the contour of local particle density in Fig.\ref{arm1}(b) and
Fig.\ref{arm2}(b), the quiver of local current density vector around
the convergence spot in p-region are plotted in Fig.\ref{armCur1}
and Fig.\ref{armCur2}, respectively. For demonstration purpose, the
local current density is plotted at every other site. The arrow on
each site denotes the local current density vector whose module and
direction are described by the size/color and orientation of the
arrow respectively. In the low energy regime (Fig.\ref{armCur1}),
the vectors of local current density converge conically to the
focusing spot [red spot in Fig.\ref{arm1}(b)]. On the other hand,
current density is converged mainly from four crossed corners in the
high energy regime (Fig.\ref{armCur2}). Furthermore, comparing
Fig.\ref{armCur1} and Fig.\ref{armCur2}, it is clear that electron
flow with larger $E_0$ gives better convergence.

\subsection {Effect of disorders in armchair nanoribbon}

As discussed in the previous sections, the clean graphene PNJ is
investigated. In a real device, the disorder is always present. In
this subsection, we study the disorder effect on the perfect
focusing in the armchair nanoribbon. We consider two kinds of
disorders, one is induced by random on-site potential
$\delta\epsilon_{\bf i}$ and the other is due to the edge
defect.\cite{edgedefect} The random on-site potentials
$\delta\epsilon_{\bf i}$ with a uniform distribution $[-w/2,w/2]$
are added near PNJ within the width of $18a$ where $w$ is disorder
strength.
%
%Since we concentrate only on the average values of the density of
%state, it is needn't to take many sample configurations. In our
%calculation, the data is obtained by averaging over $500$ disorder
%configurations.
%
The edge defect is modeled through missing atoms on the graphene
edge. We model the missing atom by setting the corresponding hopping
matrix elements to zero. The edge roughness is controlled by $p$,
the probability of a missing atom on the outermost row [the red line
in Fig.11]. For both on-site potential disorder and edge defect, all
data are obtained by averaging over $500$ configurations.

In Fig.\ref{disorder} we plot the contour of local particle density
in armchair ribbon with a sharp PNJ for $E_0=0.5t$ [same as in
Fig.\ref{arm2}(a)] in the presence of random on-site potential
disorder. In Fig.\ref{disorder}, the width of armchair ribbon is set
to $105b$, the source flow is injected from the honeycomb unit cell
at $(-210a, 0)$ and focused around the spot located at $(210a, 0)$.
Panel (a), (b), (c) and (d) correspond to different disorder
strengths $w=0$, $0.2$, $0.5$ and $1.0$. For the small random
potential strength $w$ (e.g. $w=0.2$), the interference pattern and
the focusing spot can be well kept. On the other hand, for the large
$w$ (e.g. $w=1.0$), we can see that the random potential disturbs
the interference between forward and backward scattering, so the
interference pattern is smeared, which increases the density of
state outside the focusing spot. Consequently, the intensity of
focusing spot decreases. However, we emphasize that although random
potential disturbs the interference pattern and reduces the
intensity of the focusing spot, the focused spot is clearly visible
and its size still remains unchanged. It means that the focusing
effect is robust against random potential, especially in the weak
disorder case.

In Fig.\ref{defect} we plot the contour of local particle density in
the presence of edge defect for $E_0=0.5t$. Panel (a), (b), (c) and
(d) are corresponding to the different probability $p$ of a missing
atom on the outermost row with $p=0$, $0.1$, $0.2$ and $0.5$. We
find that in the presence of edge disorder, the size of focusing
spot increases clearly and the focusing intensity is greatly
reduced. So the effect of the edge defect on the focusing effect is
more significant than that of the random potential. But the focusing
spot and interference pattern still survive and are clearly visible
(see Fig.\ref{defect}d), even in the strong edge defect case with
$p=0.5$.

To estimate the disorder strength needed to reduce the intensity of
focusing spot, in Fig.\ref{disdef}(a) and (b), we plot the maximum
value (the value at the focusing spot central $[L,0]$) of the
focused spot ${\rm LDOS}_{max}$ vs strength of random potential $w$
and the probability of a missing atom $p$. Considering the
computational cost, here we take 200 configurations and label the
error bar. From Fig.\ref{disdef}(a), we find that for weak random
potential (when $w<0.5$), ${\rm LDOS}_{max}$ hardly changes with
$w$, and focusing effect remains unchanged [see
Fig.\ref{disorder}(a), (b) and (c)]. Beyond the weak disorder regime
($w>0.5$), ${\rm LDOS}_{max}$ declines abruptly, and focusing effect
can't be kept as good as in the weak disorder regime [see
Fig.\ref{disorder}(d)]. On the other hand for the edge defect (see
Fig.\ref{disdef}(b)), we can see that the electron beam can be
focused perfectly at $p=0$ and $p=1$ because the graphene ribbon
edges are intact at both $p=0$ and $1$. When $p$ increases from $0$
to $1$, more and more atoms in edge are missing until two edges are
completely peeled. Correspondingly ${\rm LDOS}_{max}$ decreases
first and then increases since the edges are the most random when
$p$ is around 0.5. We notice in Fig.\ref{disdef}(b), comparing to
${\rm LDOS}_{max}$ near $p=1$, ${\rm LDOS}_{max}$ is reduced faster
near $p=0$. It means that the vacancy defect (a few atoms are
missing on edges) destroys focusing effect more significantly than
the adsorption defect (a few atoms are attached to edges).

\subsection{Focusing of armchair ribbon with smooth PNJ}

Up to now, we have studied focusing effect by the sharp PNJ. But in
realistic graphene based PNJ or PNPJ, the potential changes smoothly
from $E_n$ to $E_p$ within a width $d$. The width $d$ is of the
order of the separation between the graphene layer and the top gate
and $d\sim ~$tens nm.\cite{pnj} In such a smooth PNJ, backscattering
is present near PNJ in the distance proportional to $k_y$ and
interface width $d$, which reduces the possibility of Klein
tunneling. For a linear electrostatic potential $U(x)=(vk_F/d)x$,
the angular dependent transport probability\cite{Klein}
$T(\theta)=e^{-\pi(k_Fd)\sin^2(\theta)}$ where $\theta$ is incident
angle. It is obvious that the smooth PNJ will reduce the intensity
of the focused electron beam due to the decreased transport
probability $T(\theta)$. It appears that it also increases the size
of the focused spot. Actually, it is not the case due to the
following reason. For a single n-p junction (whether smooth or
sharp), the electrons (holes) with an energy equal to the chemical
potential $\mu=0$ and momentum $k_x=k_F\cos(\theta)$ transport from
conduction (valence) band to the the valence (conduction) band with
the conserved $k_y=k_F\cos(\theta)$ but $k'_x=-k_x$, leading to the
almost unchanged focusing pattern, as shown in Fig.\ref{smooth}.

The smooth PNJ can be modeled by smoothly varied Dirac energy $U(x)$
across the PNJ. In the numerical calculation, we use the following
$U(x)$,
\begin{eqnarray}
U(x)=\left\{
\begin{array}{rr}
-E_0\left[1+{{\rm sinh}(\frac{x_0}{L})}/{{\rm sinh}(\frac{x-x_0}{L})}\right], & x\leq 0 \\
 & \\
 E_0\left[1-{{\rm sinh}(\frac{x_0}{L})}/{{\rm sinh}(\frac{x+x_0}{L})}\right], & x\geq 0
\end{array}\right.
\end{eqnarray}
where $L$ is the distance between source probe located at $(-L,0) $
and PNJ at $x=0$. In Fig.\ref{smooth}(a), with $L=320 \times
0.43nm$, $U(x)$ for different $x_0$ have been plotted. In
Fig.\ref{smooth}(b, c, d) with the same parameters used in
Fig.\ref{arm1}(a) in which the sharp PNJ is used, the contour of
local particle density is re-plotted for the smooth PNJ shown in
Fig.\ref{smooth}(a). We can see that the quantum interference
between the states scattered by forward and backward scattering is
present. The density of state outside focused spot is increased.
Moreover, intensity of the convergent spot is reduced comparing to
that of sharp PNJ due to the reduced transmission probability. The
wider the PNJ interface, the more the focusing effect is reduced
because of the smaller $T(\theta)$. For example, when the PNJ width
$d=0$, the maximum of local particle density of state ${\rm
LDOS}_{max}=0.0095$ [see Fig.\ref{arm1}(a)], increasing $d$
gradually, ${\rm LDOS}_{max}=0.0068$, $0.0035$ [see
Fig.\ref{smooth}(b) and (c)]. When the PNJ width $d$ is reaching to
Fermi wavelength [such that $k_Fd\sim1$, in Fig.\ref{smooth},
$k_F\sim 1/(15a)$], the intensity of focused electron beam decrease
very slowly [see Fig.\ref{smooth}(c) and (d)]. For the very big
$d>>1/k_F$, the intensity of focused spot decreases and its size
increases continually. In addition, due to the Klein tunneling, the
focusing effect can still occur and the convergent contour is almost
the same as that of the sharp PNJ, although the intensity of
focusing spot is reduced.

\subsection{Focusing of armchair ribbon in the presence of small perpendicular magnetic field}

In the presence of small perpendicular magnetic field, the momentum
$k_y$ is not a conserved quantity. In this case, electrons and holes
are deflected in opposite directions due to the opposite Lorentz
force. So the injecting electron flow in the n-region now can't be
effectively converged in the p-region and the focusing spot is
smeared. Considering the opposite deflection for the electrons and
holes, the smeared convergent spot will move away from symmetric
point $(L,0)$. The larger the size of scattering region, the
focusing effect is less significant because of the more deflection
in the larger size. On the other hand, the transmission probability
$T$ of a single PNJ becomes magnetic field dependent on the field
scale $B^*=(\hbar/e)k_F/d$ with which the cyclotron radius
$l_{cycl}=\hbar k_F/eB$ becomes comparable to the width $d$ of PNJ.
The maximum angle rotating away from normal incidence
$\theta_{max}=\pm \arcsin(B/B^*)$. The transmission probability of a
bipolar ribbon is suppressed as $T(B<B^*) \propto
\frac{W}{d}(1-(B/B^*)^2)^{3/4}$ where $d$ is width of PNJ and $W$ is
width of ribbon.\cite{Magnet} The influence of transmission
probability on focusing effect is mainly to reduce the intensity of
focused spot. So in the presence of the weak magnetic field, not
only the intensity of focused spot is reduced but the focusing
pattern is destroyed by the deflection of electron beam as well.

In Fig.\ref{mag}, we plot the local particle density in the armchair
ribbon with small magnetic field for sharp and symmetric PNJ. For
the sharp PNJ, intensity of focused spot is reduced not so severely
as in the smooth PNJ. The magnetic field $B$ is expressed in terms
of magnetic flux $BS_0$ in the unit of $\phi_0/\pi$ where
$S_0=\frac{3}{2}\sqrt3a^2$ is the area of a honeycomb unit cell and
$\phi_0=\hbar/e$ is the flux quanta. Here $BS_0=0.0001\phi_0/\pi$
corresponds to the magnetic field $B=0.4T$. From Fig.\ref{mag}, we
can see that in the low energy regime [panel (a), $E_0=0.2t$] the
focusing effect is destroyed severely, and focusing effect is hardly
destroyed in the higher energy regime [panel (b), $E_0=0.9t$] due to
the less influence of magnetic field on electron flow with higher
energy.

\section{conclusion}
In conclusion, using the tight-binding Hamiltonian, we report the
focusing of electron flow in zigzag or armchair graphene ribbon with
a single symmetric PNJ. For a sharp PNJ, in the very low energy
regime ($|E_F-E_{n/p}|=E_0<0.05$), graphene ribbon exhibits almost
conical energy spectrum and the electron flow coming from n-region
can be converged in the p-region perfectly. When energy $E_0$
increases, however, energy spectrum gradually deviates from linear
behavior. And the band structures are different for zigzag ribbon
and armchair ribbon. For the zigzag ribbon, although the interband
scattering is symmetric with $-k^{inter}_{x,in}=k^{inter}_{x,r/t}$
but the dominant scattering, intraband scattering, is asymmetric
with $k^{intra}_{x,in} + k^{intra}_{x,r/t}\ne 0$. As a result, the
electron flow coming from source in n-region can't be converged in
the p-region. As for the armchair ribbon, only the intraband
scattering exists which is always symmetric with
$-k_{x,in}=k_{x,r/t}$ for all $k_y$. This leads to a perfect
focusing effect for all energy $E_0<t$ regardless of linear or
nonlinear dispersion relationship of Dirac Fermion. Specifically,
the electron flow converges conically in the low energy regime
($E_0<0.5t$), and converge mainly from four crossed corners in the
high energy regime ($0.5t<E_0<t$). When disorder is present, the
perfect focusing in the bipolar ribbon can be robust against
disorder induced by the random potential. The perfect focusing is
however drastically affected by the edge defect, not only the
intensity of focused spot is reduced, the size of spot is also
increased. Furthermore, when the real smooth PNJ is considered,
Klein tunneling is reduced significantly due to the backscattering.
In this case the intensity of convergent spot is reduced, but the
convergent contour still remains the same. Finally, small
perpendicular magnetic field deflects the electrons and holes in
opposite directions, which destroys the perfect focusing effect
especially in the low energy regime.

$${\bf ACKNOWLEDGMENTS}$$
We gratefully acknowledge the financial support by a RGC grant (HKU
704308P) from the Government of HKSAR and NSF-China under Grants
Nos. 10734110, 10821403, and 10974236.

\begin{figure}
\includegraphics{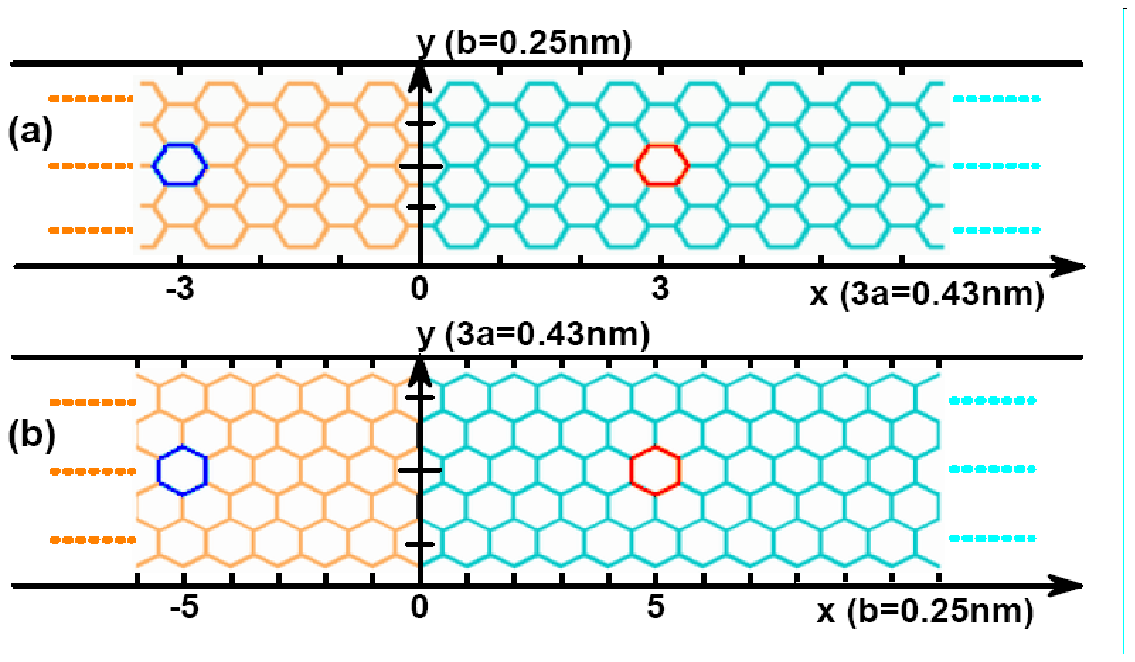}
%\includegraphics[bb=8mm 28mm 199mm 140mm,
%width=8.5cm,totalheight=5.0cm, clip=]{Fig1.eps}
\caption{ (Color
online) A schematic diagram of the graphene PNJ in a armchair ribbon
[panel (a)] and in a zigzag ribbon [panel (b)]. The graphene ribbon
is along $x$-direction and sharp p-n interface is located at $x=0$.
The electron flow injected at $(-L,0)$ [the blue area] in the
n-region [orange lattice region] is focused around the symmetric
site $(L,0)$ (the red area) in the p-region [green lattice region].}
\end{figure}
\begin{figure}
\includegraphics[bb=20mm 42mm 201mm 103mm,
width=9.0cm,totalheight=3.0cm, clip=]{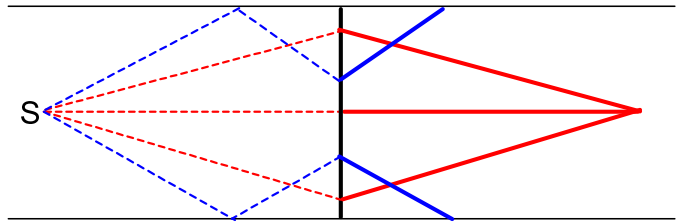} \caption{ (Color
online) After injecting from the source lead in the n-region, the
right-going electrons are scattered by the PNJ and boundary of the
bipolar graphene ribbon. }
\end{figure}
\begin{figure}
\includegraphics{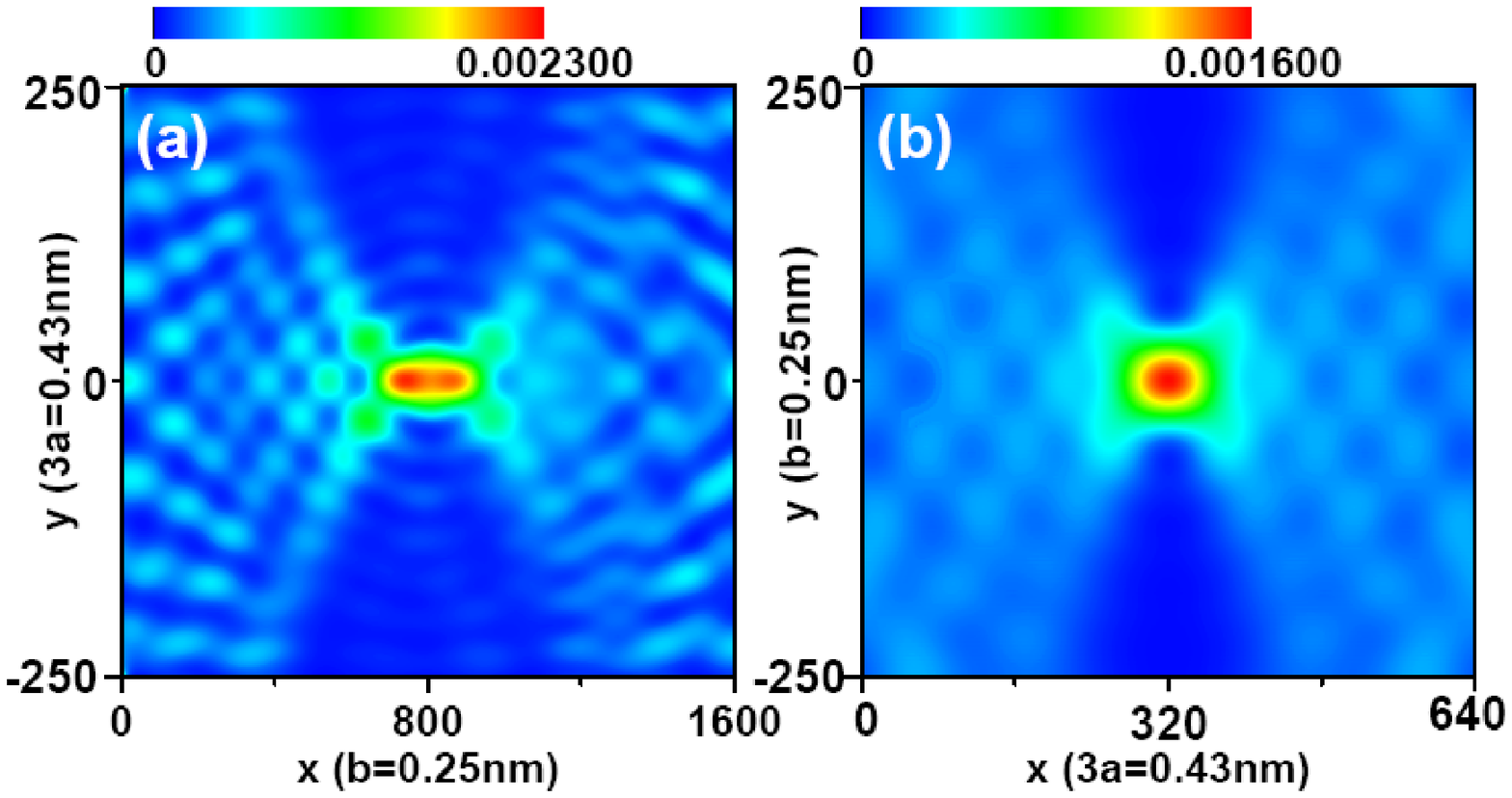}
%\includegraphics[bb=7mm 33mm 152mm 114mm,
%width=9cm,totalheight=5cm, clip=]{Graph3.eps}
\caption{ (Color
online) Distribution of local particle density $\delta
\rho(x,y)/\delta V$ in a graphene ribbon with a single sharp PNJ at
$x=0$. Panel (a): zigzag ribbon with ribbon width $W=500 \times 3a$.
The source flow is injected from the honeycomb unit cell at
$(-800b,0)$ and focused around the spot located at $(800b,0)$. Panel
(b): the armchair ribbon with ribbon width $W=501b$. The source flow
is injected from the honeycomb unit cell at $(-320\times 3a,0)$. The
other parameter used: $E_0=0.05t$.}
\end{figure}
\begin{figure}
\includegraphics[bb=-0mm 93mm 207mm 198mm,
width=11cm,totalheight=6cm, clip=]{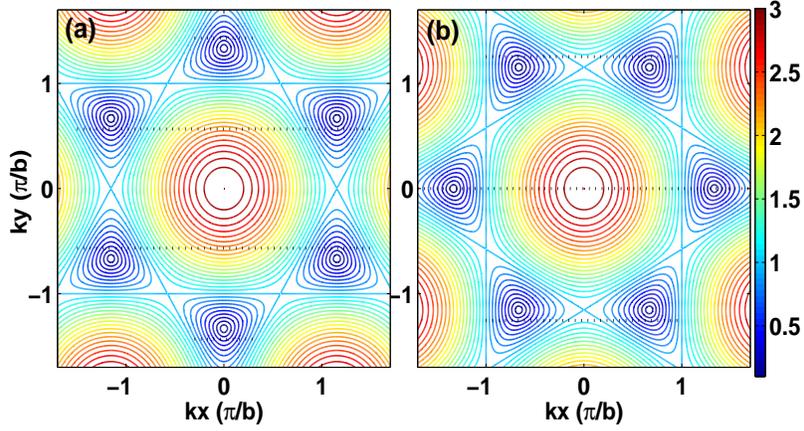} \caption{ (Color
online) The contour of dispersion relation $E(k_x,k_y)$ of graphene
sheet. The energy interval between nearest contour lines is $\delta
E=0.1t$. Panel (a): $E(k_x,k_y)$ of the graphene sheet with the
carbon-carbon bond is along the $x$-direction corresponding to
armchair ribbon or Fig.\ref{structure}(a). Panel (b): $E(k_x,k_y)$
of the graphene sheet with the carbon-carbon bond is along the
$y$-direction corresponding to zigzag ribbon or Fig.1(b).}
\end{figure}
\begin{figure}
\includegraphics[bb=11mm 11mm 190mm 147mm,
width=9cm,totalheight=6.5cm, clip=]{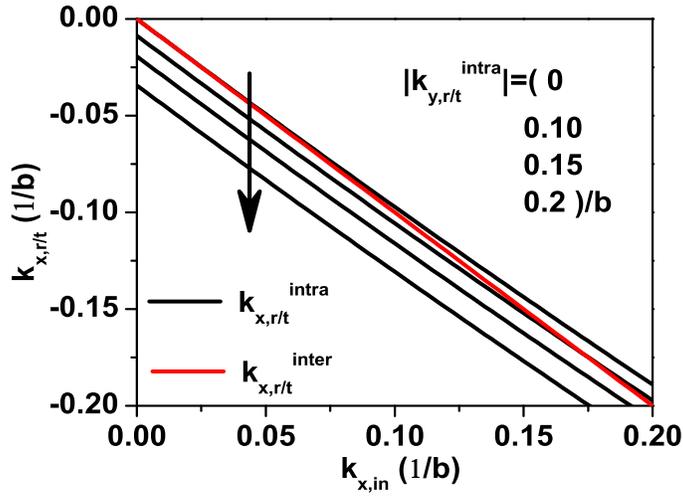} \caption{ (Color
online) Scattering momentum $k_{x,r/t}^{intra}$ and
$k_{x,r/t}^{inter}$ vs injecting momentum $k_{x,in}$ for the zigzag
ribbon with sharp PNJ. In the interband case,
$k_{x,r/t}^{inter}=-k_{x,in}$ for all conserved $k_y$ (red line).
While in the intraband case (black lines), $k_{x,r/t}^{inter}$ is
not equal to $-k_{x,in}$. Different black lines along the black
arrow are corresponding to $k_y=0$, $0.1/b$, $0.15/b$, $0.2/b$,
respectively. }
\end{figure}
\begin{figure}
\includegraphics{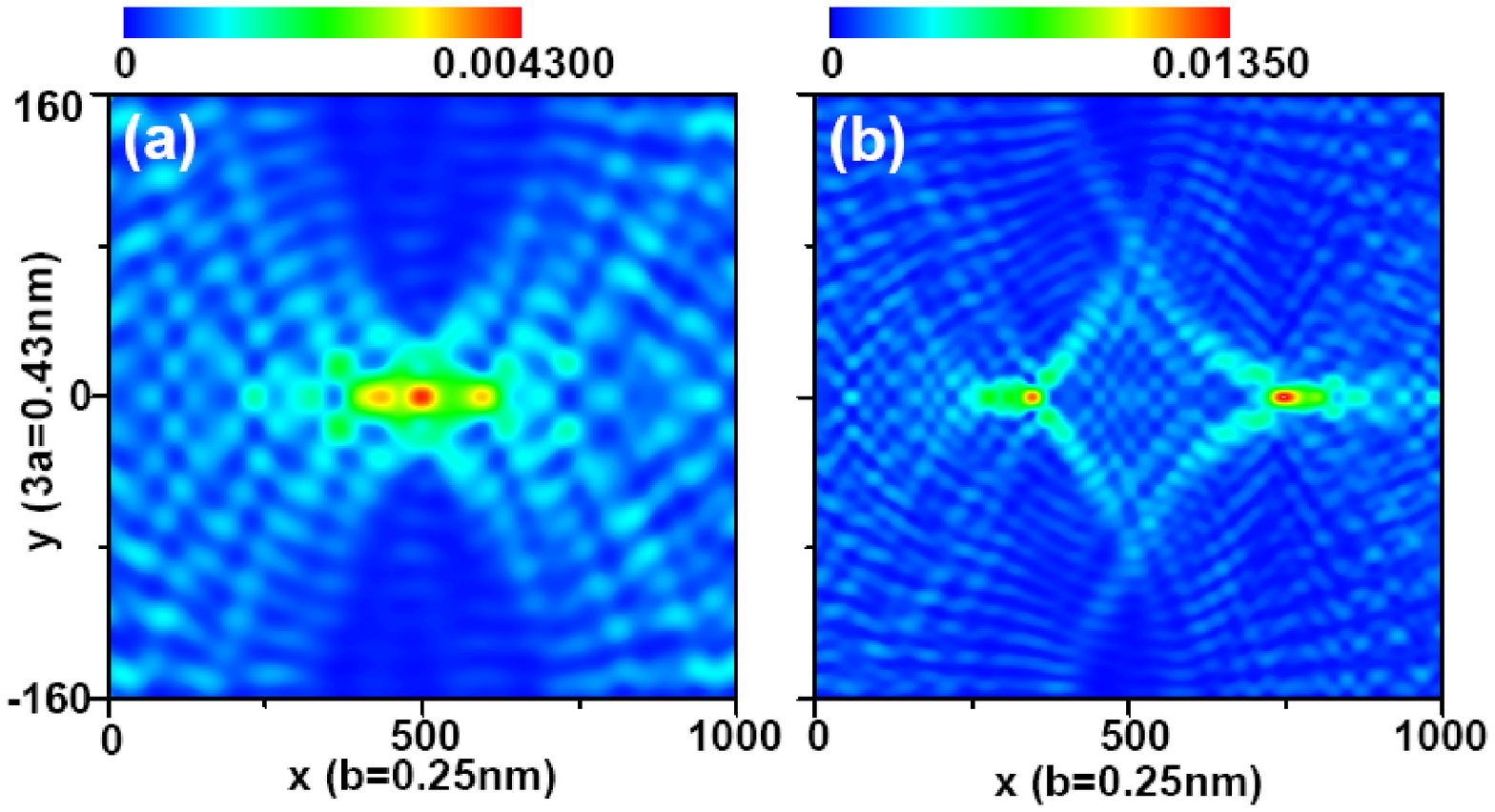}
%\includegraphics[bb=8mm 80mm 162mm 168mm,
%width=9cm,totalheight=5cm, clip=]{Graph6.eps}
\caption{ (Color
online) Contour of local particle density in zigzag ribbon with a
sharp PNJ for $E_0=0.1t$ [panel (a)] and $E_0=0.2t$ [panel (b)].}
\label{zig}
\end{figure}
\begin{figure}
\includegraphics{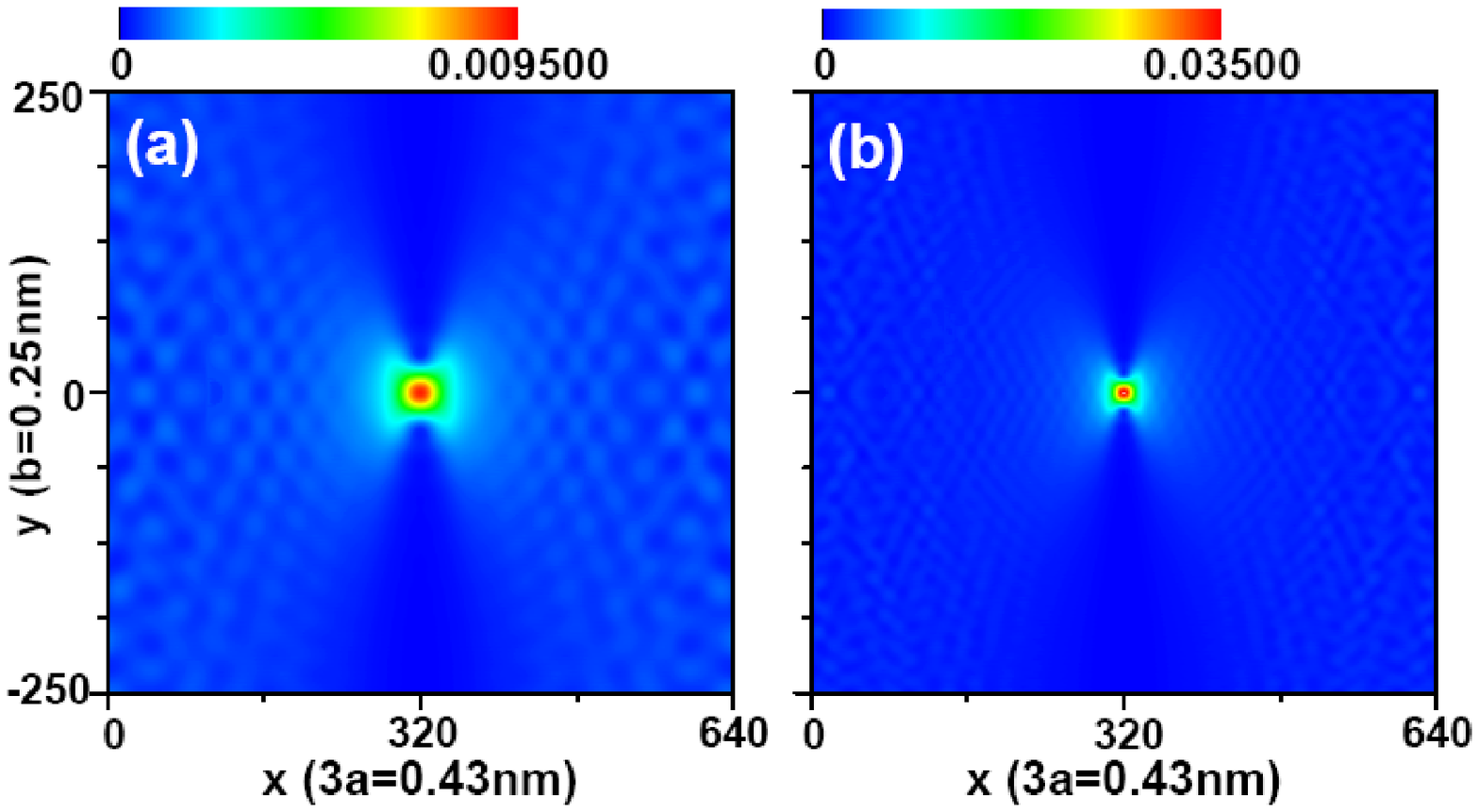}
%\includegraphics[bb=14mm 82mm 160mm 164mm,
%width=9cm,totalheight=5cm, clip=]{Graph7.eps}
\caption{ (Color
online) Contour of local particle density in armchair ribbon with a
sharp PNJ for $E_0=0.1t$ [panel (a)] and $E_0=0.2t$ [panel (b)].}
\label{arm1}
\end{figure}
\begin{figure}
\includegraphics{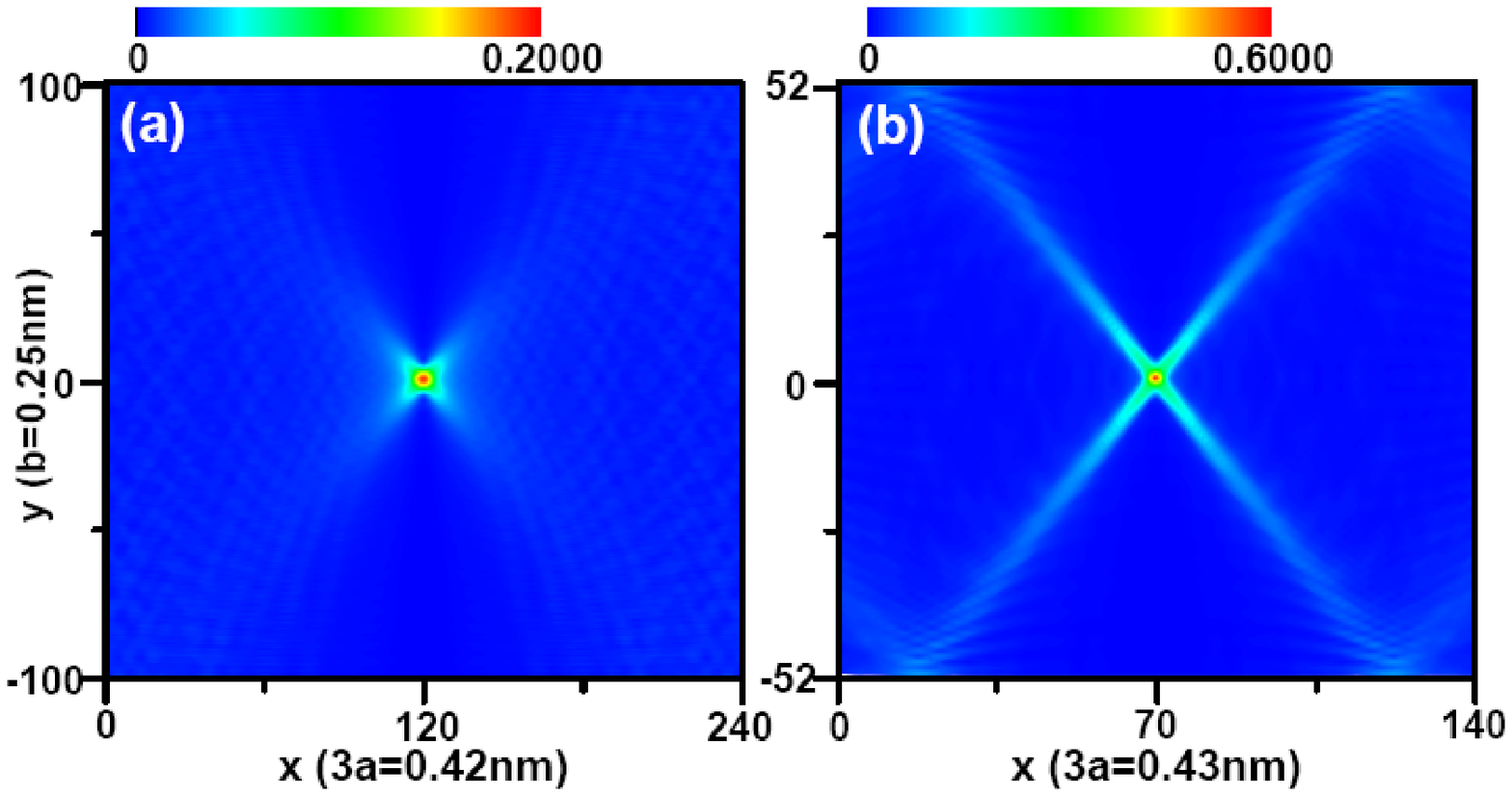}
%\includegraphics[bb=10mm 79mm 166mm 164mm,
%width=9cm,totalheight=5cm, clip=]{Graph8.eps}
\caption{ (Color
online) Contour of local particle density in armchair ribbon with a
sharp PNJ for $E_0=0.5t$ [panel (a)] and $E_0=0.9t$ [panel (b)].}
\label{arm2}
\end{figure}
\begin{figure}
\includegraphics[bb=12mm 90mm 197mm 200mm,
width=10cm,totalheight=7cm, clip=]{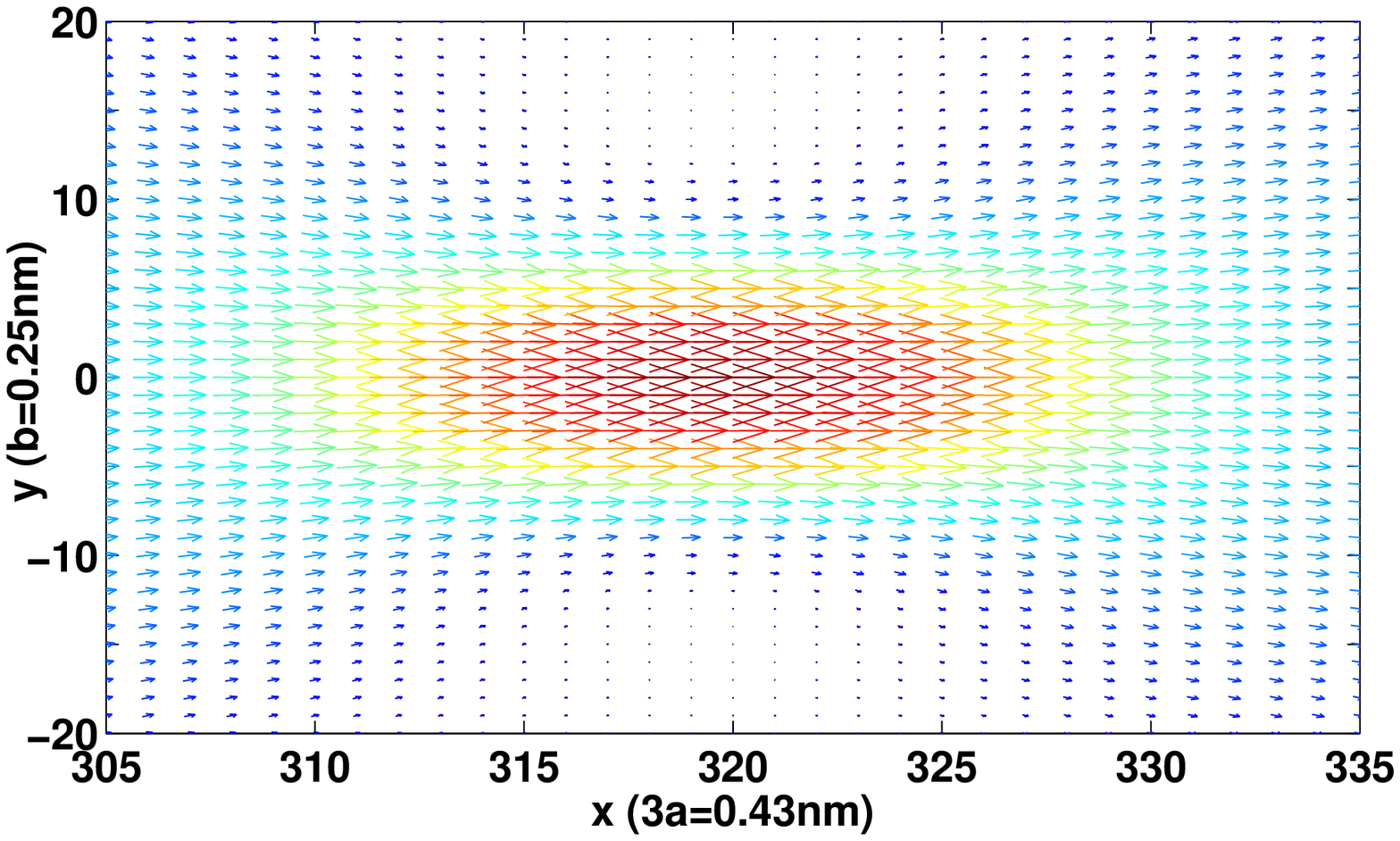} \caption{(Color online)
Instead of contour of local particle density in Fig.\ref{arm1}(b),
the quiver of local current density vector around convergence spot
is plotted.} \label{armCur1}
\end{figure}
\begin{figure}
\includegraphics[bb=1mm 90mm 197mm 200mm,
width=10cm,totalheight=7cm, clip=]{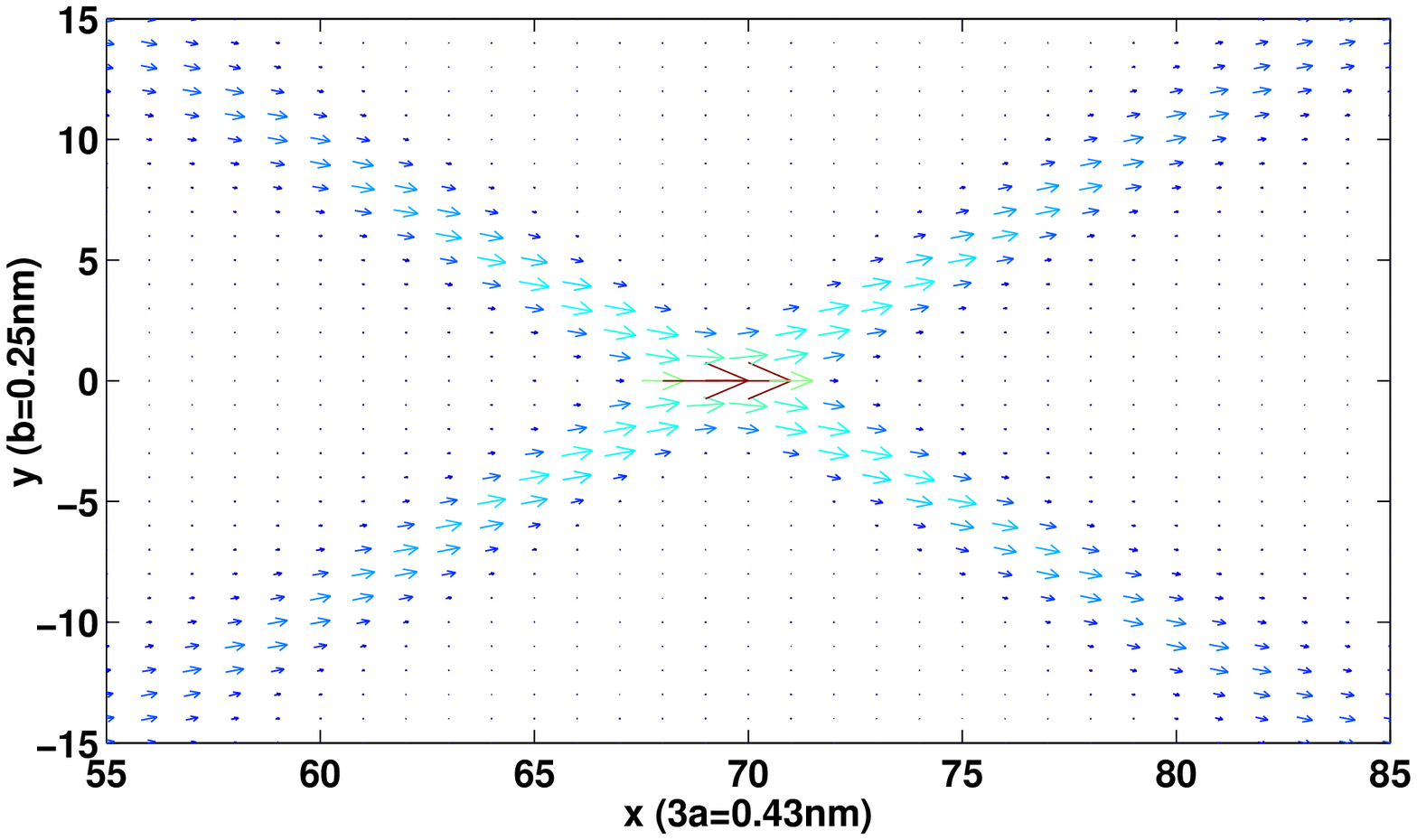} \caption{(Color
online) Instead of contour of local particle density in
Fig.\ref{arm2}(b), the quiver of local current density vector around
convergence spot is plotted.} \label{armCur2}
\end{figure}
\begin{figure}
\includegraphics[bb=10mm 11mm 210mm 117mm,
width=8cm,totalheight=4cm, clip=]{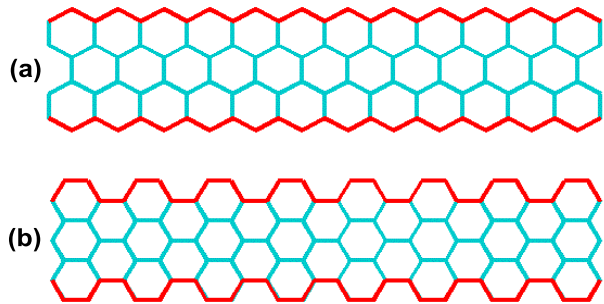} \caption{(Color Online)
panel (a): sketch of edge in the scattering region with zigzag edge.
panel (b) sketch of edge in the scattering region with armchair
edge.} \label{edge}
\end{figure}
\begin{figure}
\includegraphics{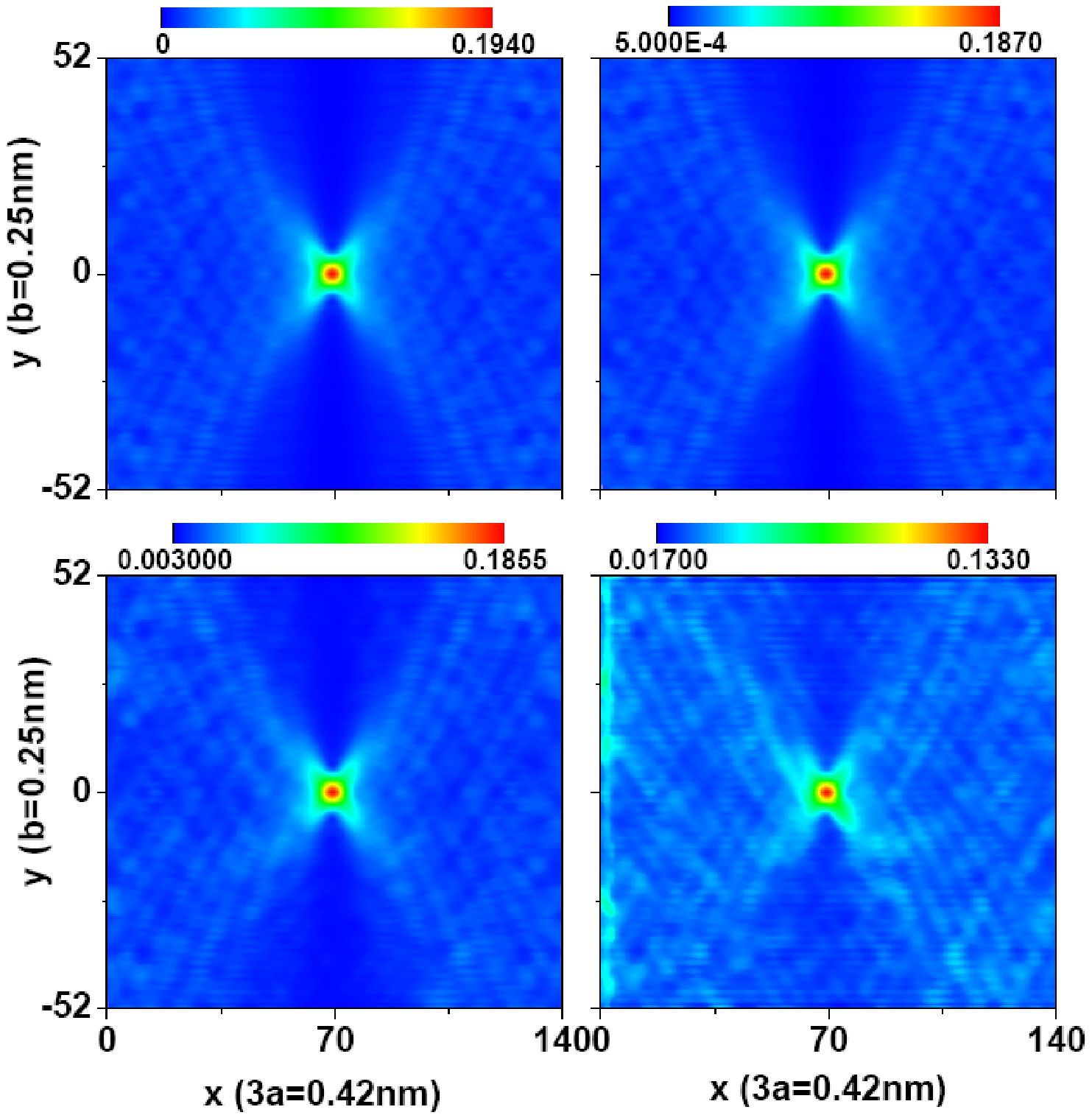}
%\includegraphics[bb=13mm 187mm 161mm 343mm,
%width=9cm,totalheight=9cm, clip=]{Graph12.eps}
\caption{(Color
Online) Contour of local particle density in the armchair ribbon
with a sharp PNJ for $E_0=0.5t$. Panel (a),  panel (b), panel (c)
and panel (d) are corresponding to random on-site potential
strengths $w=0$, $0.2$, $0.5$ and $1.0$, respectively. }
\label{disorder}
\end{figure}
\begin{figure}
\includegraphics{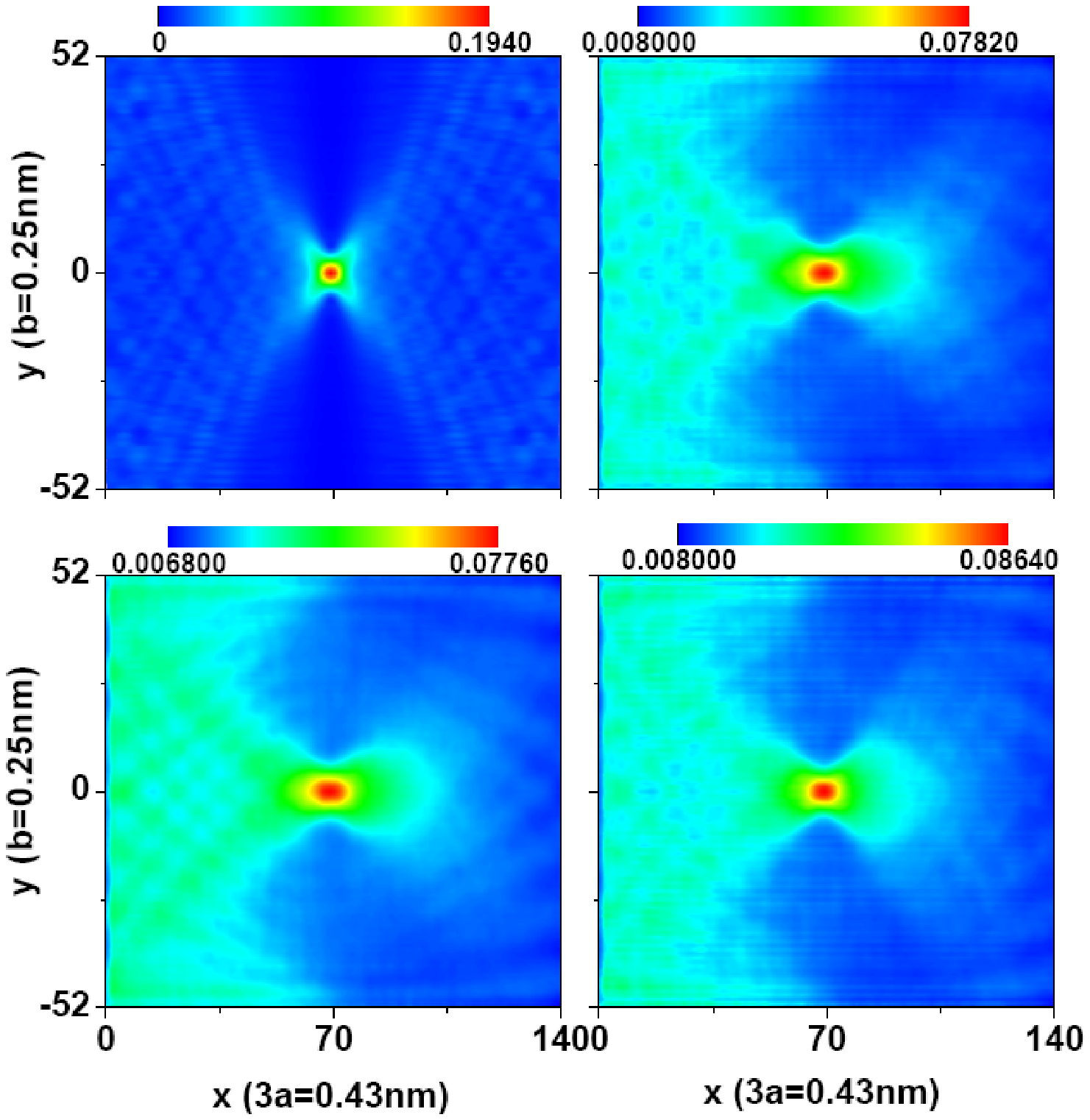}
%\includegraphics[bb=9mm 192mm 160mm 352mm,
%width=9cm,totalheight=9cm, clip=]{Graph13.eps}
\caption{(Color
Online) Contour of local particle density in armchair ribbon with a
sharp PNJ for $E_0=0.5t$. Panel (a),  panel (b), panel (c) and panel
(d) are corresponding to $p=0$, $0.1$, $0.2$ and $0.5$,
respectively.} \label{defect}
\end{figure}
\begin{figure}
\includegraphics[bb=10mm 8mm 177mm 199mm,
width=9cm,totalheight=10cm, clip=]{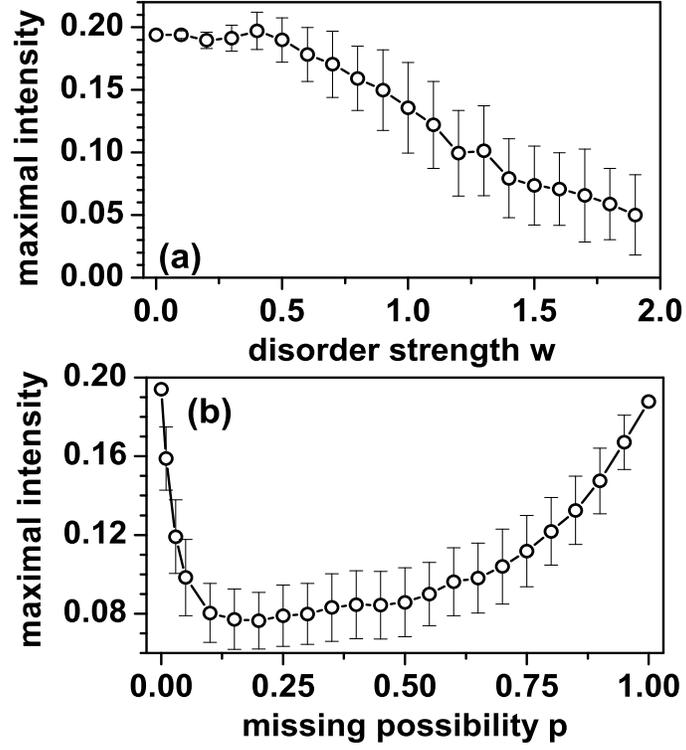} \caption{According to
Fig.\ref{disorder} and Fig.\ref{defect}, maximum value (the value at
the focusing spot central $[L,0]$) of the focusing spot ${\rm
LDOS}_{max}$ vs the strength of random potential $w$ [in panel (a)]
and probability of one missing atom $p$ [in panel (b)] are plotted,
respectively.} \label{disdef}
\end{figure}
\begin{figure}
\includegraphics{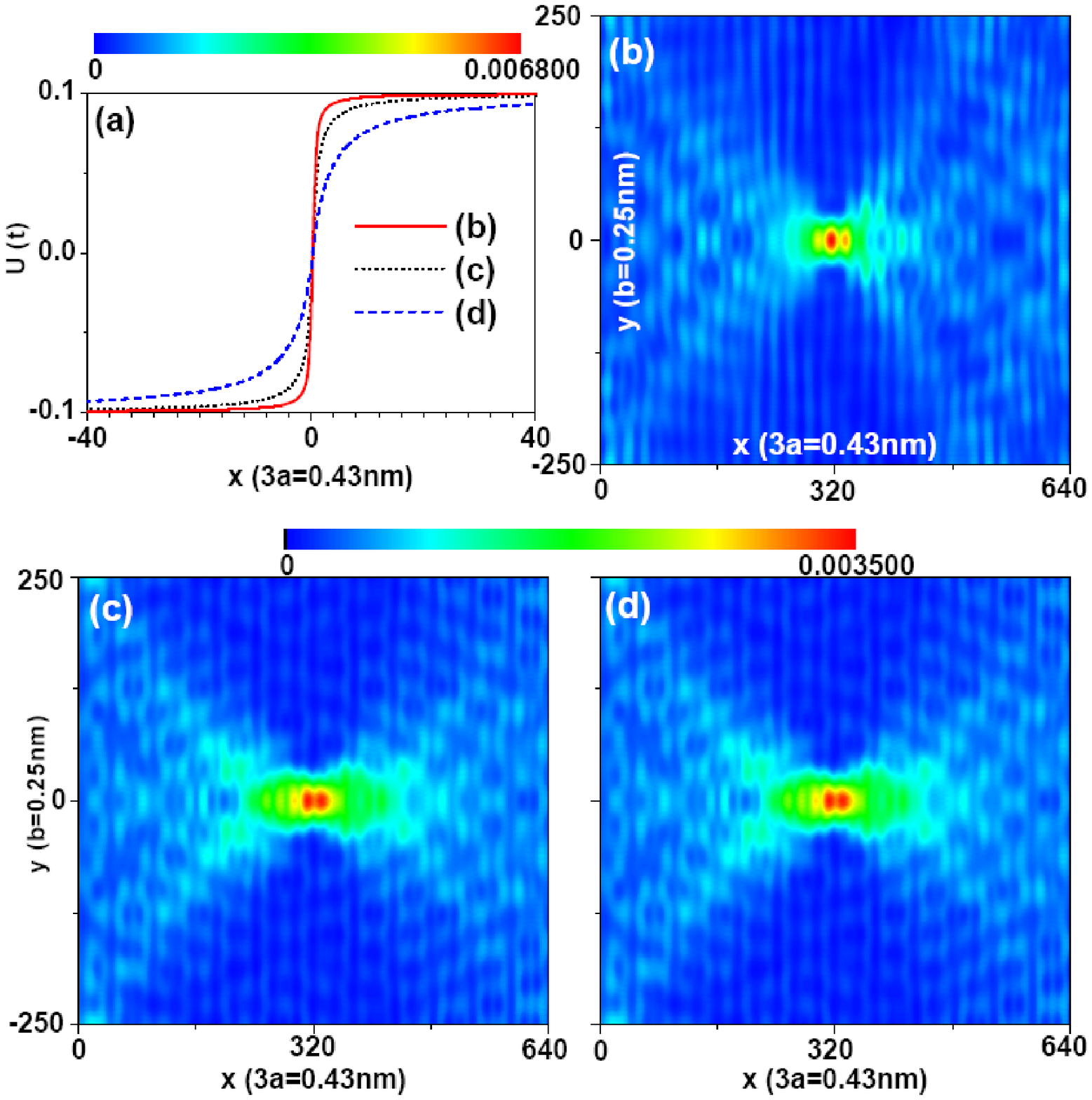}
%\includegraphics[bb=11mm 177mm 167mm 336mm,
%width=9cm,totalheight=9cm, clip=]{Graph15.eps}
\caption{(Color
Online) panel (a): smoothly changed $U(x)$ forming smooth PNJ used
in panel (b, c, d), in which $x_0=0.25*3a$, $0.75*3a$ and $3*3a$,
respectively. Panel (b, c, d): Contour of local particle density in
armchair ribbon for $E_0=0.1$ [same to Fig.\ref{arm1}(a) where sharp
PNJ is used]. For different panels, smooth PNJ shown in panel (a)
are used respectively.} \label{smooth}
\end{figure}
\begin{figure}
\includegraphics{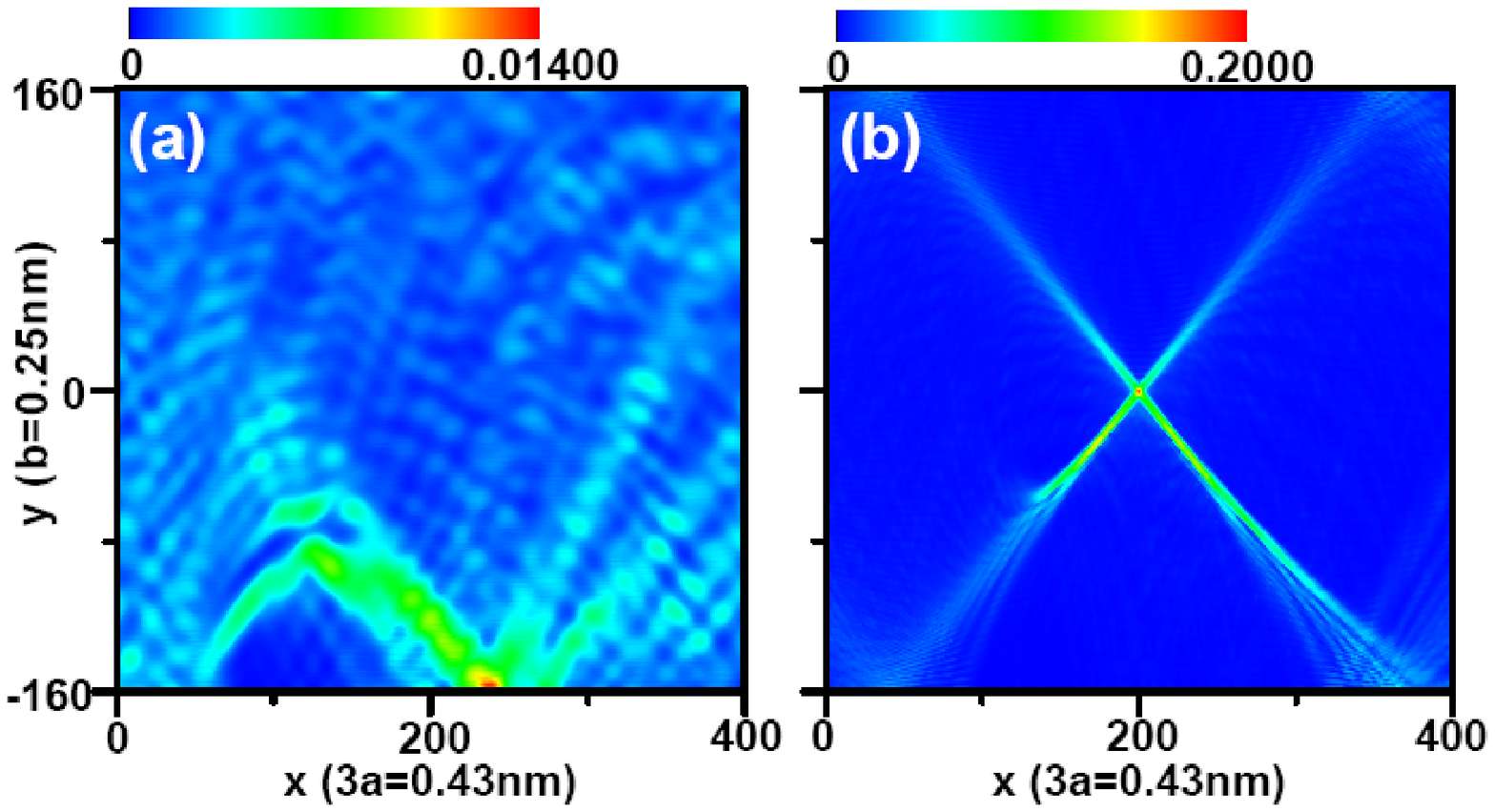}
%\includegraphics[bb=10mm 114mm 167mm 203mm,
%width=9cm,totalheight=5cm, clip=]{Graph16.eps}
\caption{ (Color online) Contour of local particle density in the
p-region for a armchair ribbon in the perpendicular weak magnetic
filed. The sharp PNJ is located at $x=0$. The magnetic field
$BS_0=0.0001\phi_0/\pi$, $E_0=0.2t$ in panel (a) and $E_0=0.9t$ in
panel (b).} \label{mag}
\end{figure}


\begin{thebibliography}{99}
\bibitem{footnote}
The band structure of a graphene ribbon is same as that of a
graphene sheet, except the zigzag edge state\cite{edgestate} would
be present in ziazag ribbon. However, for the tunneling process
beyond the first subband in conducting or valence band, the edge
state is not significant, so energy band of garphene sheet can be
safely used in graphene ribbon.
\end{thebibliography}
\end{document}